\documentclass[twocolumn]{aastex61}

\newcommand\kthr{k_\mathrm{thr}}
\newcommand\rcm{r^\mathrm{21cm}}

\submitjournal{ApJ}

\shorttitle{Stochasticity in the 21cm power spectrum}
\shortauthors{Alexander A. Kaurov}

\begin{document}

\title{Stochasticity in the 21cm power spectrum \\at the epoch of reionization and cosmic dawn}

\correspondingauthor{Alexander A. Kaurov}
\email{kaurov@ias.edu}

\author[0000-0003-0255-1204]{Alexander A. Kaurov}
\affil{Institute for Advanced Study \\
Einstein Drive, Princeton, NJ 08540, USA}

\begin{abstract}

The 21cm neutral hydrogen line is likely to be a key probe for studying the epoch of reionization and comic dawn in the forthcoming decades. This prospect stimulates the development of the theoretical basis for simulating the power spectrum of this line. Because of the beam size of the upcoming radio telescopes at high redshifts, most of the theoretical models are focused on the inhomogeneities on scales above few comoving megaparsecs. Therefore, smaller scales are often neglected and modeled with approximated sub-grid models. In this study we explore whether the perturbations on small scales ($\lesssim 1h^{-1}\mathrm{Mpc}$) can affect the 21cm signal on larger scales. Two distinct mechanism are discussed. First, we show that during the cosmic dawn small scale perturbations regulate the formation time of the first Lyman-alpha Emitters (LAE), and consequently the coupling timing of spin and kinetic temperatures. Due to the low number density of LAE, the inhomogeneity of coupling includes the shot noise and manifests itself in the observed 21cm power spectrum. Second mechanism works during the reionization when the ionized bubbles actively grow and overlap. Small scales perturbations affect the galactic properties and merger histories, and consequently the number of ionizing photons produced by each galaxy. The ionizing photons bring the perturbations from the galactic scales to the scales of ionizing fronts, affecting the 21cm power spectrum. We conclude that these two effects introduce stochasticity in the potentially observed 21cm power spectrum and, moreover, might give another perspective into the physics of the first galaxies.  

\end{abstract}


\section{Introduction}

The prospect of detection 21cm line power spectrum from high redshifts by the upcoming experiments such as HERA \citep[e.g.][]{DeBoer2016HydrogenHERA}, LOFAR \citep[e.g.][]{VanHaarlem2013}, MWA \citep[e.g.][]{Beardsley2016}, SKA \citep[e.g.][]{Mellema2012} drives the development of the theoretical models of inhomogeneous reionization. The accurate interpretation of the 21cm power spectrum of the early universe would benefit our cosmological  \citep[e.g. global optical depth, ][]{Liu2016EliminatingCosmology} and astrophysical constraints \citep[e.g. the temperature of the intergalactic medium, ][]{Pober2015PAPER-64MEDIUM}.

The models of the inhomogeneous cosmic reionization can be roughly subdivided into two groups. First one is the semi-analytical/semi-numerical models, many of which are based on the analytical model by \citet{Furlanetto2004} and its various extensions that include additional physics \citep[i.e. ][]{Battaglia2013, Kaurov2013a, Kaurov2014, Sobacchi2014}. Models based on this approach effectively approximate the radiation transfer with the excursion set approach \citep[for instance, 21CMFAST by ][]{Mesinger201121cmfast:Signal}. Second group is the numerical simulations with radiation transfer and other physics explicitly modeled up to some physical scale \citep{Ciardi2003SimulatingReionization, McQuinn2007TheReionization,Iliev2009, Aubert2010, Petkova2010SimulationsFeedback, Friedrich2011, Shapiro2012, So2014, Gnedin2014, Duffy2014Low-massReionization, Aubert2015EMMA:Transfer, Ocvirk2016CosmicUniverse, Pawlik2016TheResults}.

Due to the physics complexity and the range of physical scales involved during the epoch of reionization \citep{Pritchard2012}, any method has some approximations and assumptions. In the numerical simulations those are ``hidden'' in the sub-grid models. In the analytical methods the approximations are incorporated into various parameters, that are often not well defined; for instance, the escape fraction of ionizing photons that is assumed to be a simple function of halo mass. Therefore, numerically or analytically, the physical processes are considered up to some smallest scale -- the size of the grid cell. It makes both approaches deterministic, i.e. they output a fixed reionization history for a given set of initial conditions (ICs) and physics. In other words, the ionization history of an individual cell is fully determined by the initial conditions of the whole box. In this study we explore whether there are effects that could be overlooked because of this assumption.

The size of the smallest cell is often motivated by the expected beam size of the planned 21cm telescopes, which is order of a few $h^{-1}$Mpc at $z>6$. Thus, we investigate how smaller scales (order of 100s of $h^{-1}$kpc) may affect the observable 21cm power spectrum on $\gtrsim 1h^{-1}$Mpc scales. To do so we adopt a simulation (\S\ref{sec:methods}) that does resolve smaller scales and galaxy formation. By perturbing the ICs of the simulation on small scales (above some wavenumber $\kthr$) and rerunning the simulation, we measure the effect on the 21cm power spectrum. For an external observer, who sees 21cm fluctuations only below $\kthr$, the effect looks like a stochastic term, which increases the total amplitude of the power spectrum. In \S\ref{sec:results} we discuss two distinct signatures that manifest themselves during the cosmic dawn and the epoch of reionization. Finally, we discuss the limitations of our approach, and the theoretical and observational prospects of these effects in \S\ref{sec:discus}. 

\section{Methods}
\label{sec:methods}

\begin{figure*}
\centering
\includegraphics[scale=0.9]{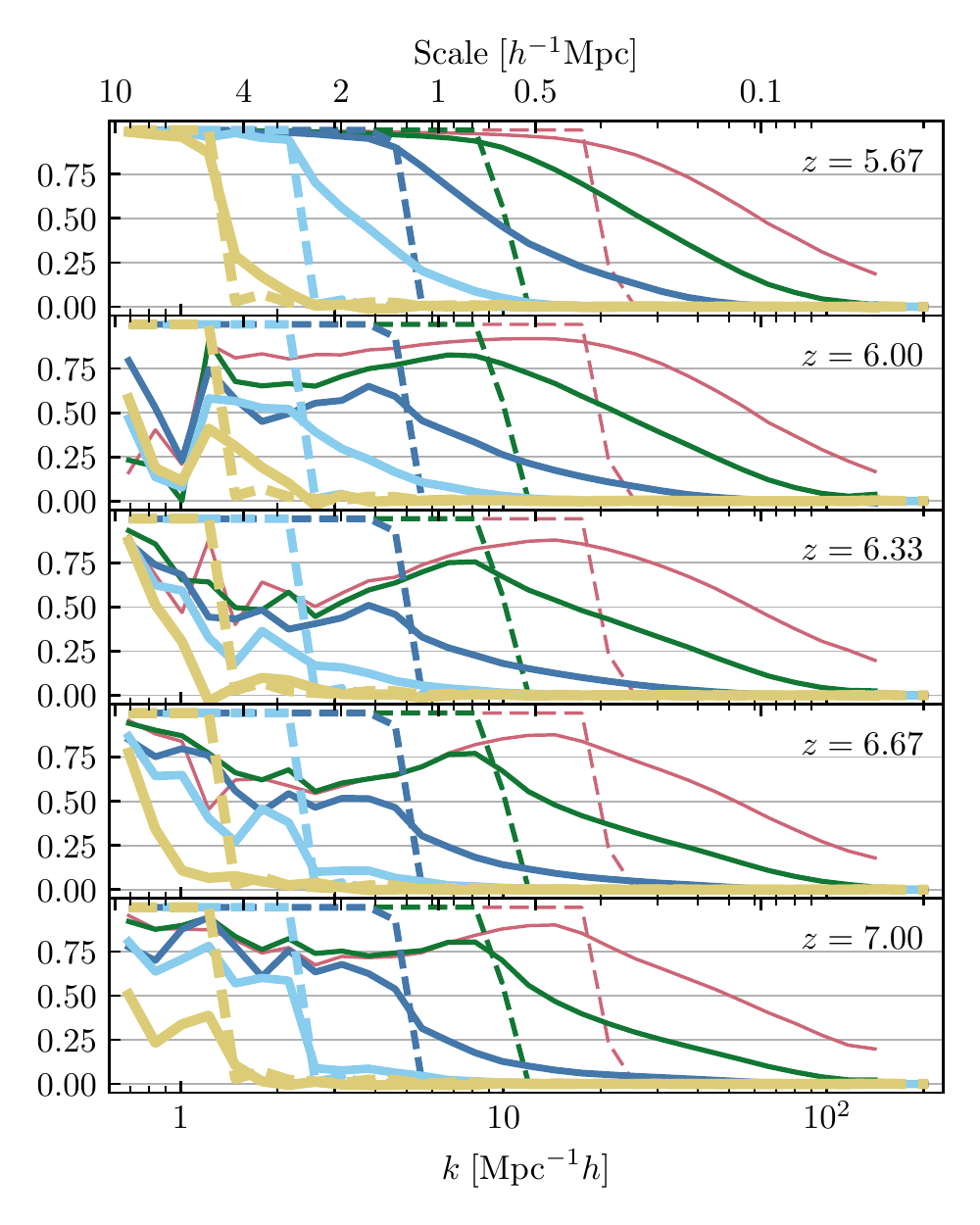}
\includegraphics[scale=0.9]{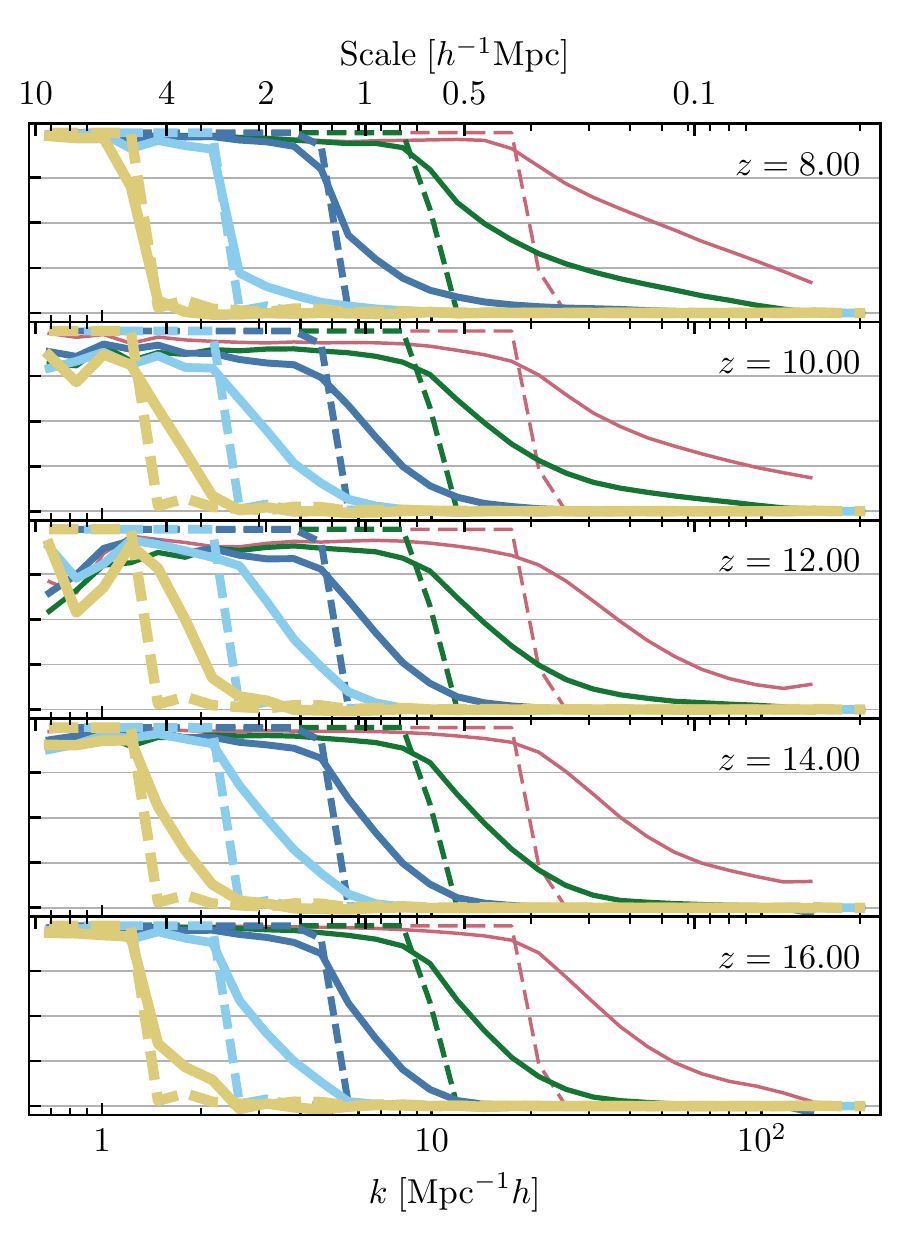}
\caption{\label{fig:rk_21cm_all}
Power spectrum correlation coefficient of the 21cm brightness temperature, $\rcm$, that spans across the redshifts of the cosmic dawn (\textit{right column}) and the epoch of reionization (\textit{left column}). Color and thickness corresponds to the different values of $\kthr$ and the dashed lines show its value. The dashed lines can also be interpreted as the cross correlation of the ICs, $r^\mathrm{IC}$. The divergence of $\rcm$ from unity at $k<\kthr$ corresponds to the stochastic effect described in \S\ref{sec:results}.}
\end{figure*}

\subsection{Numerical simulations}

We adopt the numerical simulation used in the Cosmic Reionization On Computers (CROC) project \citep{Gnedin2014, 2014ApJ...793...30G}. The features of the code that important specifically for this study are: adaptive mesh refinement (that allows to model star formation within galaxies and resolve scales up to 100 pc at $z\sim6$), radiative transfer, Ly$\alpha$ coupling and X-ray heating. The general properties of the 21cm power spectrum of this particular simulations is studied in \citet{Kaurov2015a}. 

In this study we need to use tens of simulations; therefore, we adopt a relatively small size for the simulation boxes -- 10 $h^{-1}$Mpc. The initial condition (IC) for a such box is defined on a $256^3$ grid. We randomly generate one IC that we further refer to as ``reference'' IC. Then, we generate ``perturbed'' ICs by randomly changing the angles and amplitudes (obviously, preserving the correct matter power spectrum) for the Fourier modes of the reference IC above some threshold wave number, $\kthr$. 

We adopted the following values for $\kthr$ -- (1.26, 2.51, 5.03, 10.05, 20.11) $h\mathrm{Mpc}^{-1}$ that corresponds to a half, a quarter, etc. of the box's size. For each $\kthr$ we generate 8 random realizations. The example of a perturbed ICs can be seen in the first row of Figure \ref{fig:eor}.

Then, all boxes are evolved with the exactly same physics. The starting redshift is 50. Thus, the epoch of cosmic dawn and the coupling of the kinetic and spin temperatures of the neutral hydrogen is covered by the simulation.

The coupling mostly goes through the Ly$\alpha$ radiation -- Wouthuysen-–Field effect \citep{Wouthuysen1952OnLine.,Field1959TheHydrogen.,Field1959TheProfile.}. The Ly$\alpha$ radiation is not propagated with the radiation transfer; it is assumed that the Ly$\alpha$ background has effectively infinite mean free path.

The preheating of the IGM with X-rays is also calculated; however, for this study it is not important. In the adopted model the heating kicks relatively late and does not affect the effects discussed in this paper. In \S\ref{sec:cd} we discuss what would be the consequences if the early heating does take place.

\begin{figure*}
\includegraphics[width=\textwidth]{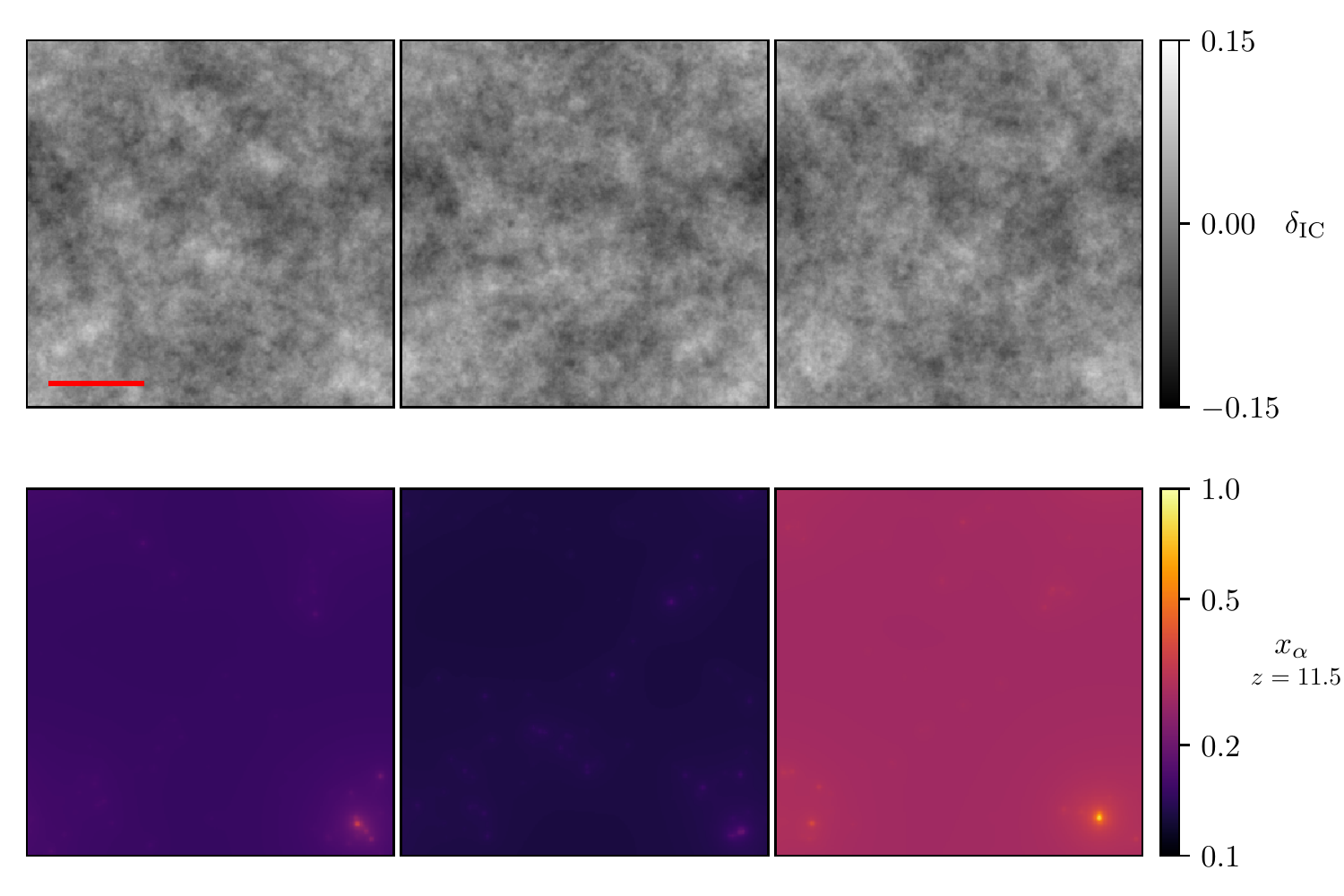}
\caption{\label{fig:cd}The panels show volume weighted mean projections of the $10\,h^{-1}\mathrm{Mpc}$ simulation boxes for $\kthr\sim2.5\,h\mathrm{Mpc}^{-1}$. The rows represent: the initial conditions and the coupling coefficient $x_\alpha$ at $z\sim7$. The leftmost column corresponds to the ``reference'' simulation, and the other two are random perturbed ICs and corresponding simulated $x_\alpha$ fields. The scale of perturbations is displayed as red segment in the top-left panel.}
\end{figure*}

\subsection{Comparison}

We perform the comparison between different simulation boxes using the cross correlation coefficient:

\begin{equation}
r^f_{ij}(k) = P^f_{ij}(k) / \sqrt{P^f_{ii}(k)\,P^f_{jj}(k)}
\end{equation}
where $P^f_ij$ is the cross power spectrum between simulations $i$ and $j$ for filed $f$. For a given value of $\kthr$ we have 9 simulations (the reference one and 8 perturbed), and we calculate $r^f_{ij}(k)$ for all pairs. Then, for each set we calculate the averaged $r$ among all pairs of simulations in order to reduce noise:
\begin{equation}
r^f(k) = \langle r^f_{ij}(k) \rangle_{ij}.
\end{equation}

Since all of the fields we consider (excluding the ICs) are extremely non-Gaussian, the statistics based on power spectrum is more illustrative, rather than quantitative.

Another way of thinking about $r^f(k)$ is that it shows the ratio between the averaged power spectrum and the power spectrum of the average, i.e.:
\begin{equation}
r^f(k) = \frac{P^{\langle f_{i}\rangle_i}(k)}{\langle P^f_{ii}(k) \rangle_i}.
\end{equation}
This definition might be more intuitive. Imagine, we have a deterministic model that works on scales below $\kthr$ and is capable of predicting the ensemble averaged field $f$ for all ICs perturbed at above $\kthr$. Then, this model will generate the power spectrum $\langle P^f_{ii}(k) \rangle_i$. However, the ``true'' expected power spectrum for a field with known ICs below $\kthr$ is $P^{\langle f_{i}\rangle_i}(k)$. Therefore, $1-r^f(k)$ can be interpret as a fraction of the power spectrum missed by the deterministic model.

%
%
%
%

\section{Results}
\label{sec:results}

In our main Figure \ref{fig:rk_21cm_all} the cross correlation coefficient of the 21cm brightness power spectrum $r^\mathrm{21cm}(k)$ is plotted for a range of redshifts and $\kthr$. The behavior of the $r^\mathrm{21cm}(k)$ is easily explainable at redshifts $\sim16$, $\sim8$, and $5.67$, where $\rcm$ diverges from unity at the scale of $k_{thr}$. At those redshifts the universe is either fully neutral or ionized, and the intensity of 21cm line simply traces the density field.

However, at other redshifts there is a noticeable divergence that have stochastic behavior. We call it ``stochastic'' since small scale perturbations affect larger scales. Thus, an observer that resolves scales only down to, for example, $k\sim10$, will see the fluctuation at $k\sim1$ as random.   

We further discuss the physical nature of divergence in the next two subsections -- the cosmic dawn (CD) and the epoch of reionziation (EoR),  -- since the physical effects that cause the stochasticity are different.

\subsection{Cosmic Dawn: $10 \lesssim z \lesssim 15$}
\label{sec:cd}

\begin{figure*}
\includegraphics[width=\columnwidth]{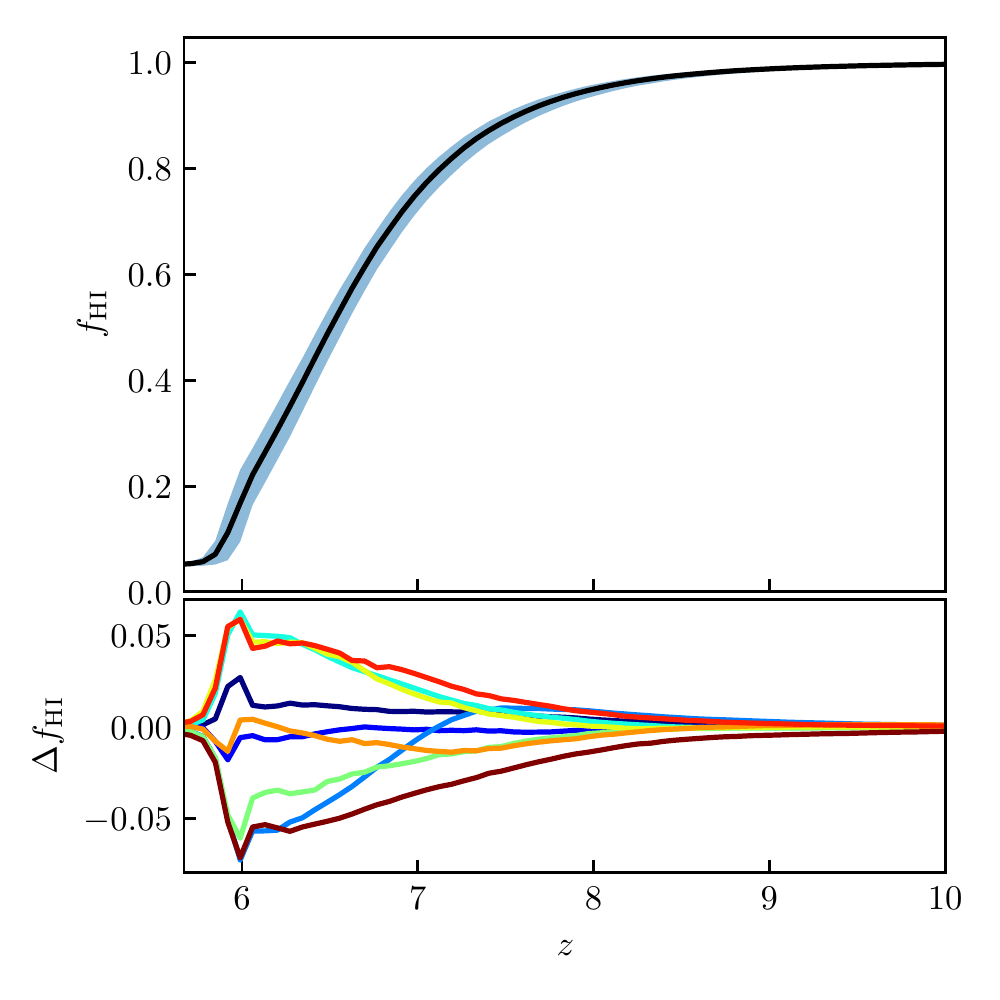}
\includegraphics[width=\columnwidth]{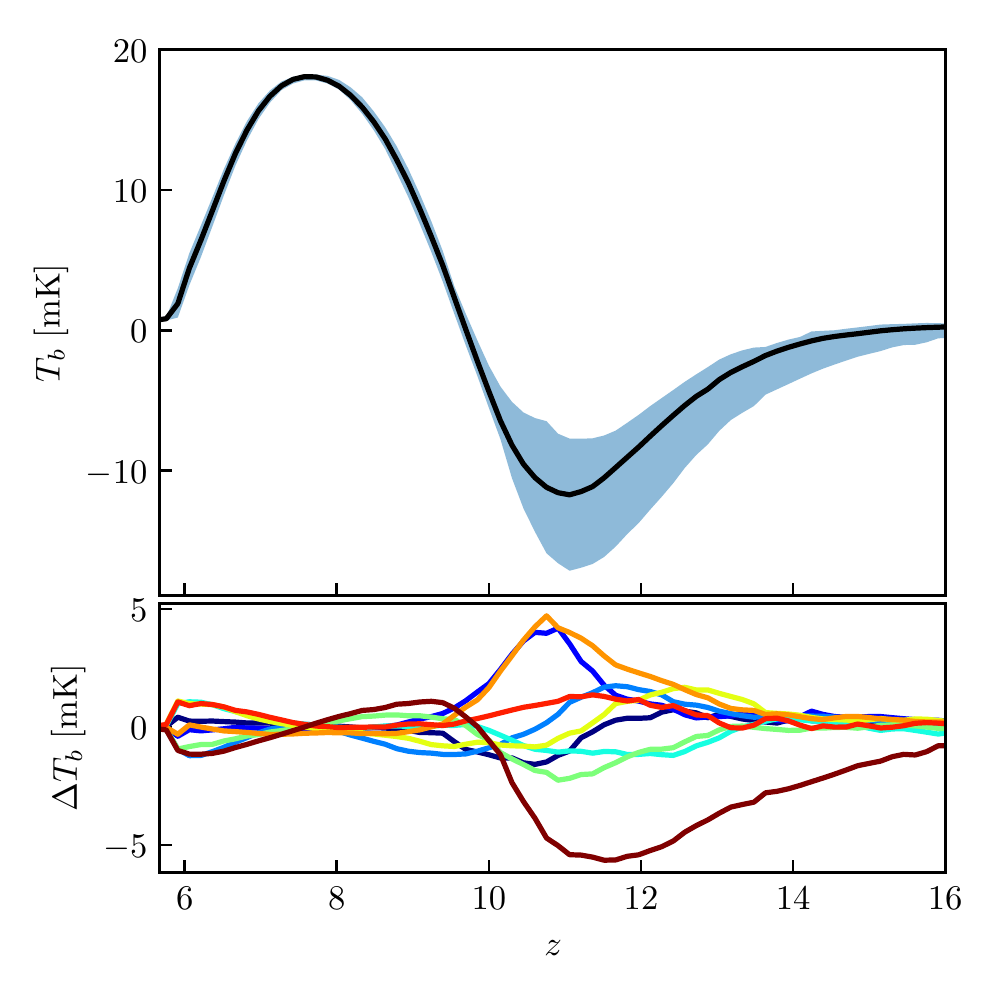}
\caption{\label{fig:21hist}The reionization history for the $\kthr \sim 2.5h\mathrm{Mpc}^{-1}$ set of simulations \textit{(left panel)}, and the global 21cm brightness temperature (\textit{right panel}). Shaded regions correspond to the total scatter among 9 simulations, bottom panels show the deviations of individual simulations. The reference simulation corresponds to the dark blue color.}
\end{figure*}

During the CD the universe remains mostly neutral. It is the time when the spin temperature of neutral hydrogen couples with the kinetic temperature of gas. The biased distribution of the Lyman Alpha Emitters (LAE) leads to the inhomogeneity in coupling and, consequently, affects the 21cm power spectrum (as well as absolute mean spin temperature).

The signatures caused by these inhomogeneities can give clues regarding the nature of the first emitters \citep{Ghara201521Velocities,Fialkov2015ReconstructingSignal}. What we observe in our numerical experiment is the stochasticity in the formation of these first sources. Since the coupling is achieved relatively quickly, the total number density of sources that takes part in generating sufficient Ly$\alpha$ background is low. The randomness in the timing of their formation and in their intensity cause the effect that can be interpreted as \textit{shot noise}. The right column in Figure \ref{fig:rk_21cm_all} shows that the contribution from the shot noise can reach up to 20\% at $z\sim 10-12$. 

In the Figure \ref{fig:cd} we show the reference and two perturbed ICs ($\kthr\sim2.5\,h\mathrm{Mpc}^{-1}$) and corresponding coupling coefficient, $x_\alpha$ \citep[i.e.][]{Pritchard2012}. We chose to show volume weighted projection instead of a slice, since the number of LAE is low, and a slice can coincide only with one of them.

It can be seen that the ICs differ, but preserve global overdensity trends on larger scales. The position of the LAE is approximately the same; however, their intensity is different as well as total Ly$\alpha$ background. Meanwhile, the differences in the total ionized fraction in all cases at $z\sim11.5$ is negligible. 

The differences in the gas temperature is also negligible in our case, since it takes place later on due to the chosen stellar population model. Ultimately, the early heating by, for instance X-ray binaries, may have a similar effect. In recent years many such models were studied \citep{Madau199721Redshift,Venkatesan2001HeatingBackground,Oh2001ReionizationClusters,Madau2004EarlyMiniquasars,Ricotti2004X-rayMeasurement,Furlanetto2006TheRedshifts,Pritchard200721-cmReionization,Ciardi2010LyMedium,Mirabel2011StellarUniverse,Mesinger2013SignaturesUniverse,Tanaka2016TheRadiation,Madau2017RADIATIONBINARIES}; however, as of right now, due to the absence of the observable data, nothing definite can be said regarding the IGM preheating. 

In the Figure \ref{fig:21hist} the ionization history and the global (within the box) 21cm brightness temperature for $\kthr\sim2.5\,h\mathrm{Mpc}^{-1}$ are presented. The ionization history does not change much; therefore, the global optical depth is effectively not sensitive to the perturbations on these scales. All boxes completely ionize at $z \sim 6$ partially because there is uniform QSO contribution to the ionizing background implemented in the code. Meanwhile, the total brightness temperature in the box fluctuates significantly. This fact shows that our box is too small for properly studying such a sparse abundance of LAE. 


\subsection{The epoch of reionization: $6 \lesssim z \lesssim 8$}
\label{sec:eor}

\begin{figure*}
\includegraphics[width=\textwidth]{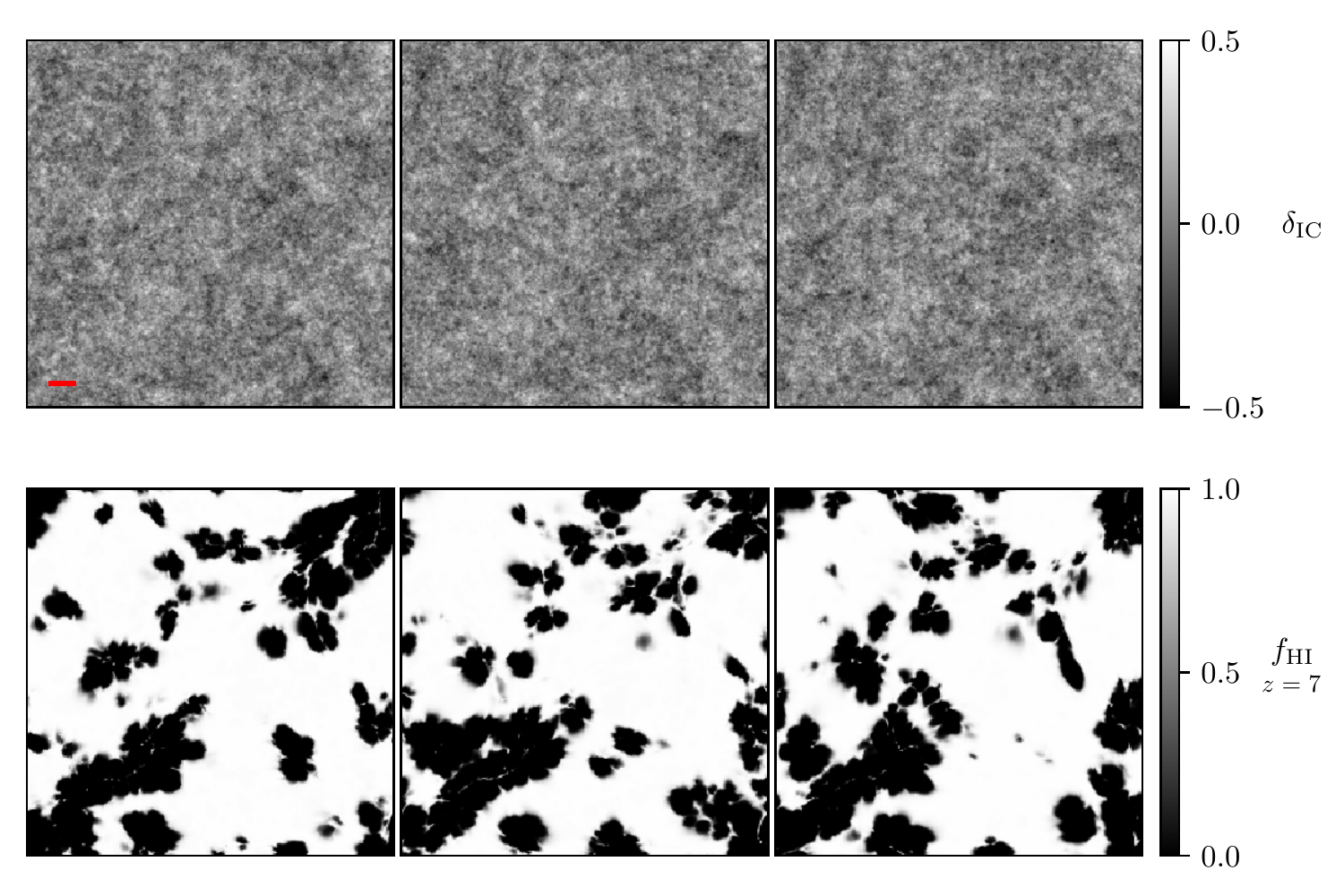}
\caption{\label{fig:eor}The panels show three simulation boxes for $\kthr \sim 10h\mathrm{Mpc}^{-1}$. The rows represent the initial conditions and the ionization field at $z\sim7$. The leftmost column correspond to the ``reference'' simulation, and the other two are random perturbed ICs and corresponding simulated ionization fields. The scale of perturbations is displayed as red segment in the top-left panel. It can be seen by eye that the ionization fields have differences on much larger scales.}
\end{figure*}

\begin{figure}
\includegraphics[scale=0.9]{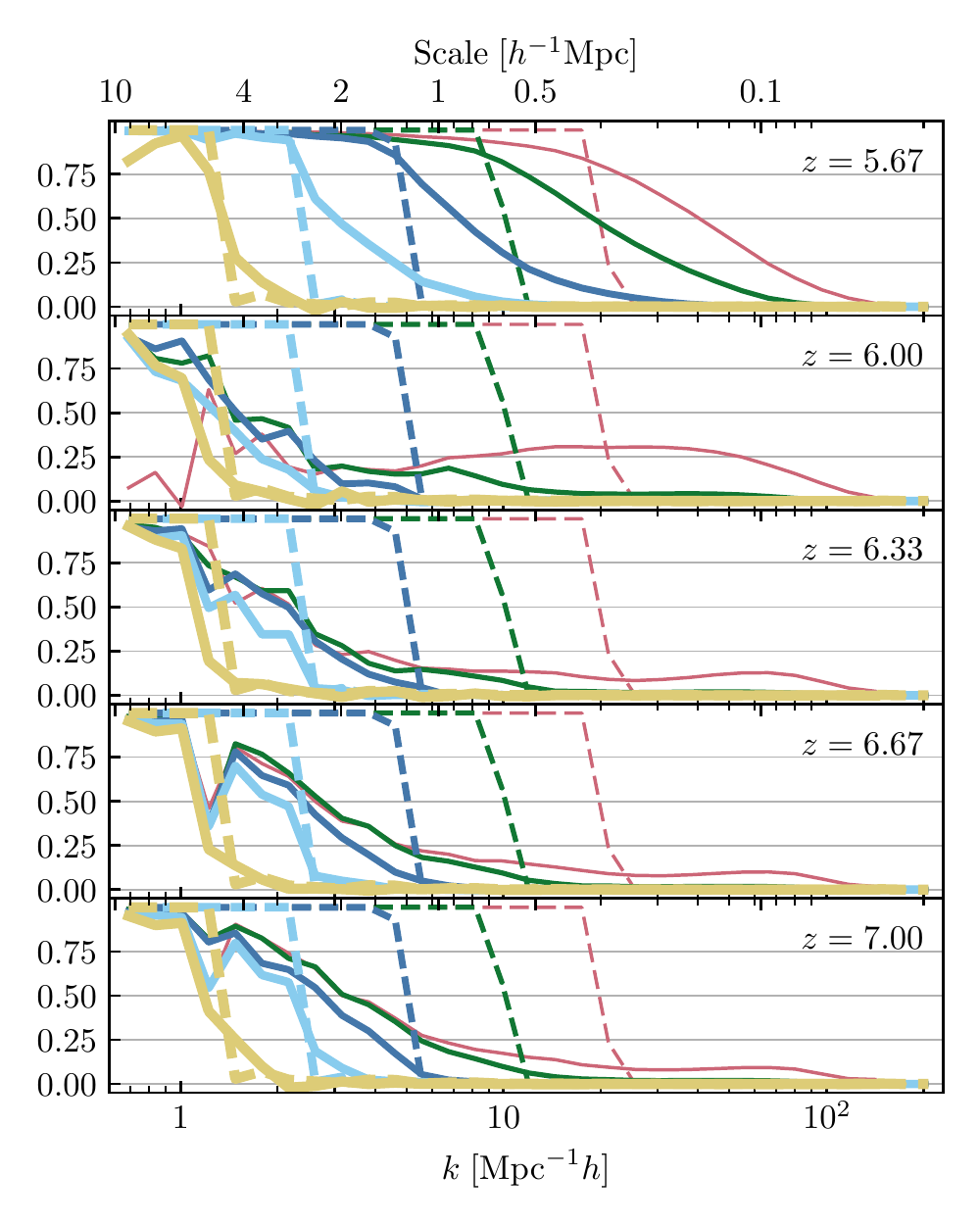}
\caption{\label{fig:ionr}The cross correlation coefficient for the neutral fraction, $r^{f_\mathrm{HI}}(k)$, during the epoch of cosmic reionization. The notation is identical to the Figure \ref{fig:rk_21cm_all}.}
\end{figure}


During the epoch of reionization the spin temperature is already tightly coupled to the kinetic gas temperature. Therefore, the power spectrum signal is regulated mainly by two things -- the distribution of gas and the morphology of the ionized bubbles. At the beginning and at the of the EoR the power spectrum follows the density perturbations; therefore, $r^\mathrm{21cm}$ is unity below $k_{thr}$ and gradually drops at higher wavenumber  (see left column in Figure \ref{fig:rk_21cm_all}). In between these two regimes the inhomogeneity of the ionization in the IGM influences the power spectrum. In result, we see that $r$ diverges from one even below $k_{thr}$, and the effect can reach up to 50\%.

The Figure \ref{fig:eor} shows the slices from three ICs and resulted ionization field morphology approximately at the middle of the reionization. The ionization field exhibits visually detectable differences at scales above the scale of perturbations.


In the Figure \ref{fig:ionr} the cross correlation of the neutral hydrogen fraction is plotted. It shows that divergence is even greater. This is because it is sensitive to the shape of ionizing fronts only, while the 21cm power spectrum consists from  two components: the neutral regions of IGM and the semi-neutral filaments and Lyman Limit Systems inside ionized bubbles \citep[see][for a detailed discussion of these two components]{Kaurov2015a}. Since the 21cm power spectrum at higher wavenumber is mostly defined by the filaments, the cross correlation coefficient ``recovers'' at higher wavenumber.

These differences in the morphology of the ionization fronts are caused by the peculiar histories of ionizing photons production of individual galaxies. The escape fraction of a galaxy is likely to be very spherically inhomogeneous and fluctuating in time \citep[e.g.][]{Gnedin2007, Trebitsch2017FluctuatingGalaxies}. These fluctuations are originally seeded in the smallest scales of the ICs. Somewhat larger scales are responsible for altering the merging histories of galaxies. All these effects combined cause the changes in the morphology of ionizing bubbles, and consequently alter the 21cm power spectrum.

We do not expect here to see the shot noise similar to one described in \S\ref{sec:cd} because the number of sources that participate in ionizing is much greater compared to those driving initial coupling. 


\,

\section{Discussion}
\label{sec:discus}

We have shown how the stochastic component can contribute order of 50\%/20\% at the EoR/CD of the total 21cm power spectrum at $k \sim 1 \mathrm{Mpc}^{-1}h$ (Figure \ref{fig:rk_21cm_all}). However, we have done it only for a particular simulation and chosen set of parameters; therefore, we emphasize that the results of our study have mostly qualitative value. In order to make quantitative predictions, one has to perform a proper parameter study across different star and galaxy formation models, which probably significantly affects the amplitudes of described effects.

In addition, there are numerical limitations caused by the computational cost of running dozens of simulations. In order to avoid them, one should come up with a more numerically effective approach. Firstly, we adopted a single ``reference'' IC. This particular IC might be an outlier, and be not representative\footnote{After the study was completed, we realized that the strategy of using only one reference realization of the IC probably was not the best choice. It would be more efficient to have 4 pairs of the ICs with same realizations below $\kthr$ and randomized above $\kthr$, instead of one reference and 8 perturbed ICs.}. Secondly, the size of the box is not sufficient for studying the effect of the shot noise in Ly$\alpha$ coupling, since the shot noise is still dominant on the level of the box itself. Also, during the EoR our box size is sufficient only for probing the regime when bubbles are smaller $10h^{-1}\mathrm{Mpc}$, i.e. only the earliest stages. Finally, the number of perturbed ICs is low. With only 8 random perturbations we can probe only $1\sigma$. The effects of the shot noise are likely to be non-Gaussian.

From modeling perspective, we see two possible ways to account for the stochasticity effects reported here (once they are properly estimated). For the analytical methods it might be useful first to decompose the power spectrum into the signal from the neutral IGM and semi-neutral objects within ionized bubbles \citep{Kaurov2015a}. Then, introduce the correction for the former one, since according to our results it should be dominant. In the numerical methods the stochasticity can be incorporated into sub-grid models or into the recipe of galaxy formation as a random component.

In \cite{Kaurov2016} we made an attempt to fit a semi-analytical model (based on the excursion set formalism) into the numerical simulation (same as used in this paper). There we essentially tried to fit a deterministic model into a model with the stochastic component. We managed to do it to some extent; however, given the results of this paper we think that a fitting procedure needs to be improved in order to account for the stochasticity separately. 

As for observational perspective, the described effects (due to their randomness) can only increase the amplitude of the 21cm power spectrum, and, therefore, boost its detectability. Moreover, it would be especially interesting to decompose the observed power spectrum into deterministic and stochastic component. If the models of the CD will become accurate enough to extract the shot noise contribution, it would give a hint regarding the abundance and formation timing of the first LAE.

\bibliographystyle{yahapj}
\bibliography{Mendeley.bib}

\begin{thebibliography}{}
\providecommand\natexlab[1]{#1}
\providecommand\JournalTitle[1]{#1}

\bibitem[{Aubert {et~al.}(2015)Aubert, Deparis, Ocvirk, Aubert, Deparis, \&
  Ocvirk}]{Aubert2015EMMA:Transfer}
Aubert, D., Deparis, N., Ocvirk, P., {et~al.} 2015,
  \href{http://dx.doi.org/10.1093/mnras/stv1896}{\JournalTitle{Monthly Notices
  of the Royal Astronomical Society}, 454, 1012}

\bibitem[{Aubert \& Teyssier(2010)}]{Aubert2010}
Aubert, D., \& Teyssier, R. 2010,
  \href{http://dx.doi.org/10.1088/0004-637X/724/1/244}{\JournalTitle{The
  Astrophysical Journal}, 724, 244}

\bibitem[{Battaglia {et~al.}(2013)Battaglia, Trac, Cen, \&
  Loeb}]{Battaglia2013}
Battaglia, N., Trac, H., Cen, R., \& Loeb, a. 2013,
  \href{http://dx.doi.org/10.1088/0004-637X/776/2/81}{\JournalTitle{The
  Astrophysical Journal}, 776, 81}

\bibitem[{Beardsley {et~al.}(2016)Beardsley, Hazelton, Sullivan, Carroll,
  Barry, Rahimi, Pindor, Trott, Line, Jacobs, Morales, Pober, Bernardi, Bowman,
  Busch, Briggs, Cappallo, Corey, de~Oliveira-Costa, Dillon, Emrich,
  Ewall-Wice, Feng, Gaensler, Goeke, Greenhill, Hewitt, Hurley-Walker,
  Johnston-Hollitt, Kaplan, Kasper, Kim, Kratzenberg, Lenc, Loeb, Lonsdale,
  Lynch, McKinley, McWhirter, Mitchell, Morgan, Neben, Thyagarajan, Oberoi,
  Offringa, Ord, Paul, Prabu, Procopio, Riding, Rogers, Roshi, Shankar, Sethi,
  Srivani, Subrahmanyan, Tegmark, Tingay, Waterson, Wayth, Webster, Whitney,
  Williams, Williams, Wu, \& Wyithe}]{Beardsley2016}
Beardsley, A.~P., Hazelton, B.~J., Sullivan, I.~S., {et~al.} 2016,
  \href{http://dx.doi.org/10.3847/1538-4357/833/1/102}{\JournalTitle{The
  Astrophysical Journal, Volume 833, Issue 1, article id. 102, 19 pp. (2016).},
  833}

\bibitem[{Ciardi {et~al.}(2010)Ciardi, Salvaterra, Di~Matteo, Ciardi,
  Salvaterra, \& Di~Matteo}]{Ciardi2010LyMedium}
Ciardi, B., Salvaterra, R., Di~Matteo, T., {et~al.} 2010,
  \href{http://dx.doi.org/10.1111/j.1365-2966.2009.15843.x}{\JournalTitle{Monthly
  Notices of the Royal Astronomical Society}, 401, 2635}

\bibitem[{Ciardi {et~al.}(2003)Ciardi, Stoehr, White, Ciardi, Stoehr, \&
  White}]{Ciardi2003SimulatingReionization}
Ciardi, B., Stoehr, F., White, S. D.~M., {et~al.} 2003,
  \href{http://dx.doi.org/10.1046/j.1365-8711.2003.06797.x}{\JournalTitle{Monthly
  Notices of the Royal Astronomical Society}, 343, 1101}

\bibitem[{DeBoer {et~al.}(2016)DeBoer, Parsons, Aguirre, Alexander, Ali,
  Beardsley, Bernardi, Bowman, Bradley, Carilli, Cheng, Acedo, Dillon,
  Ewall-Wice, Fadana, Fagnoni, Fritz, Furlanetto, Glendenning, Greig,
  Grobbelaar, Hazelton, Hewitt, Hickish, Jacobs, Julius, Kariseb, Kohn,
  Lekalake, Liu, Loots, MacMahon, Malan, Malgas, Maree, Mathison, Matsetela,
  Mesinger, Morales, Neben, Patra, Pieterse, Pober, Razavi-Ghods, Ringuette,
  Robnett, Rosie, Sell, Smith, Syce, Tegmark, Thyagarajan, Williams, \&
  Zheng}]{DeBoer2016HydrogenHERA}
DeBoer, D.~R., Parsons, A.~R., Aguirre, J.~E., {et~al.} 2016,
  \href{http://arxiv.org/abs/1606.07473
  http://www.arxiv.org/pdf/1606.07473.pdf}{\JournalTitle{arXiv:1606.07473
  [astro-ph]}}

\bibitem[{Duffy {et~al.}(2014)Duffy, Wyithe, Mutch, \&
  Poole}]{Duffy2014Low-massReionization}
Duffy, A.~R., Wyithe, J. S.~B., Mutch, S.~J., \& Poole, G.~B. 2014,
  \href{http://dx.doi.org/10.1093/mnras/stu1328}{\JournalTitle{Monthly Notices
  of the Royal Astronomical Society, Volume 443, Issue 4, p.3435-3443}, 443,
  3435}

\bibitem[{Fialkov {et~al.}(2015)Fialkov, Barkana, Cohen, Fialkov, Barkana, \&
  Cohen}]{Fialkov2015ReconstructingSignal}
Fialkov, A., Barkana, R., Cohen, A., {et~al.} 2015,
  \href{http://dx.doi.org/10.1103/PhysRevLett.114.101303}{\JournalTitle{Physical
  Review Letters}, 114, 101303}

\bibitem[{Field(1959{\natexlab{a}})}]{Field1959TheHydrogen.}
Field, G.~B. 1959{\natexlab{a}},
  \href{http://dx.doi.org/10.1086/146653}{\JournalTitle{The Astrophysical
  Journal}, 129, 536}

\bibitem[{Field(1959{\natexlab{b}})}]{Field1959TheProfile.}
---. 1959{\natexlab{b}},
  \href{http://dx.doi.org/10.1086/146654}{\JournalTitle{The Astrophysical
  Journal}, 129, 551}

\bibitem[{Friedrich {et~al.}(2010)Friedrich, Mellema, Alvarez, Shapiro, \&
  Iliev}]{Friedrich2011}
Friedrich, M.~M., Mellema, G., Alvarez, M.~a., Shapiro, P.~R., \& Iliev, I.~T.
  2010,
  \href{http://dx.doi.org/10.1111/j.1365-2966.2011.18219.x}{\JournalTitle{Arxiv
  preprint}, 19, 19}

\bibitem[{Furlanetto {et~al.}(2004)Furlanetto, Zaldarriaga, \&
  Hernquist}]{Furlanetto2004}
Furlanetto, S., Zaldarriaga, M., \& Hernquist, L. 2004,
  \href{http://dx.doi.org/10.1086/423025}{\JournalTitle{ApJ}, 613, 1}

\bibitem[{Furlanetto(2006)}]{Furlanetto2006TheRedshifts}
Furlanetto, S.~R. 2006,
  \href{http://dx.doi.org/10.1111/j.1365-2966.2006.10725.x}{\JournalTitle{Monthly
  Notices of the Royal Astronomical Society}, 371, 867}

\bibitem[{Ghara {et~al.}(2015)Ghara, Choudhury, \&
  Datta}]{Ghara201521Velocities}
Ghara, R., Choudhury, T.~R., \& Datta, K.~K. 2015,
  \href{http://dx.doi.org/10.1093/mnras/stu2512}{\JournalTitle{Monthly Notices
  of the Royal Astronomical Society}, 447, 1806}

\bibitem[{Gnedin(2014)}]{Gnedin2014}
Gnedin, N.~Y. 2014,
  \href{http://dx.doi.org/10.1088/0004-637X/793/1/29}{\JournalTitle{The
  Astrophysical Journal}, 793, 29}

\bibitem[{Gnedin \& Kaurov(2014)}]{2014ApJ...793...30G}
Gnedin, N.~Y., \& Kaurov, A.~A. 2014,
  \href{http://dx.doi.org/10.1088/0004-637X/793/1/30}{\JournalTitle{The
  Astrophysical Journal}, 793, 30}

\bibitem[{Gnedin {et~al.}(2007)Gnedin, Kravtsov, \& Chen}]{Gnedin2007}
Gnedin, N.~Y., Kravtsov, A.~V., \& Chen, H.-W. 2007,
  \href{http://dx.doi.org/10.1086/524007}{\JournalTitle{The Astrophysical
  Journal, Volume 672, Issue 2, article id. 765-775, pp. (2008).}, 672}

\bibitem[{Iliev {et~al.}(2008)Iliev, Pen, McDonald, Shapiro, Mellema, \&
  Alvarez}]{Iliev2009}
Iliev, I.~T., Pen, U.-L., McDonald, P., {et~al.} 2008,
  \href{http://dx.doi.org/10.1007/s10509-008-9865-9}{\JournalTitle{Astrophysics
  and Space Science}, 320, 145}

\bibitem[{Kaurov(2016)}]{Kaurov2016}
Kaurov, A.~A. 2016,
  \href{http://dx.doi.org/10.3847/0004-637X/831/2/198}{\JournalTitle{The
  Astrophysical Journal}, 831, 198}

\bibitem[{Kaurov \& Gnedin(2013)}]{Kaurov2013a}
Kaurov, A.~A., \& Gnedin, N.~Y. 2013,
  \href{http://dx.doi.org/10.1088/0004-637X/771/1/35}{\JournalTitle{The
  Astrophysical Journal}, 771, 35}

\bibitem[{Kaurov \& Gnedin(2014)}]{Kaurov2014}
---. 2014,
  \href{http://dx.doi.org/10.1088/0004-637X/787/2/146}{\JournalTitle{The
  Astrophysical Journal}, 787, 146}

\bibitem[{Kaurov \& Gnedin(2016)}]{Kaurov2015a}
---. 2016,
  \href{http://dx.doi.org/10.3847/0004-637X/824/2/114}{\JournalTitle{The
  Astrophysical Journal}, 824, 114}

\bibitem[{Liu {et~al.}(2016)Liu, Pritchard, Allison, Parsons, Seljak, \&
  Sherwin}]{Liu2016EliminatingCosmology}
Liu, A., Pritchard, J.~R., Allison, R., {et~al.} 2016,
  \href{http://dx.doi.org/10.1103/PhysRevD.93.043013}{\JournalTitle{Physical
  Review D}, 93, 043013}

\bibitem[{Madau \& Fragos(2017)}]{Madau2017RADIATIONBINARIES}
Madau, P., \& Fragos, T. 2017,
  \href{https://arxiv.org/pdf/1606.07887.pdf}{\JournalTitle{arXiv:1606.07887v2}}

\bibitem[{Madau {et~al.}(1997)Madau, Meiksin, Rees, Madau, Meiksin, \&
  Rees}]{Madau199721Redshift}
Madau, P., Meiksin, A., Rees, M.~J., {et~al.} 1997,
  \href{http://dx.doi.org/10.1086/303549}{\JournalTitle{The Astrophysical
  Journal}, 475, 429}

\bibitem[{Madau {et~al.}(2004)Madau, Rees, Volonteri, Haardt, Oh, Madau, Rees,
  Volonteri, Haardt, \& Oh}]{Madau2004EarlyMiniquasars}
Madau, P., Rees, M.~J., Volonteri, M., {et~al.} 2004,
  \href{http://dx.doi.org/10.1086/381935}{\JournalTitle{The Astrophysical
  Journal}, 604, 484}

\bibitem[{McQuinn {et~al.}(2007)McQuinn, Lidz, Zahn, Dutta, Hernquist,
  Zaldarriaga, McQuinn, Lidz, Zahn, Dutta, Hernquist, \&
  Zaldarriaga}]{McQuinn2007TheReionization}
McQuinn, M., Lidz, A., Zahn, O., {et~al.} 2007,
  \href{http://dx.doi.org/10.1111/j.1365-2966.2007.11489.x}{\JournalTitle{Monthly
  Notices of the Royal Astronomical Society}, 377, 1043}

\bibitem[{Mellema {et~al.}(2012)Mellema, Koopmans, Abdalla, Bernardi, Ciardi,
  Daiboo, de~Bruyn, Datta, Falcke, Ferrara, Iliev, Iocco, Jeli{\'{c}}, Jensen,
  Joseph, Kloeckner, Labroupoulos, Meiksin, Mesinger, Offringa, Pandey,
  Pritchard, Santos, Schwarz, Semelin, Vedantham, Yatawatta, \&
  Zaroubi}]{Mellema2012}
Mellema, G., Koopmans, L., Abdalla, F., {et~al.} 2012,
  \href{http://dx.doi.org/10.1007/s10686-013-9334-5}{\JournalTitle{Experimental
  Astronomy, Volume 36, Issue 1-2, pp. 235-318}, 36, 235}

\bibitem[{Mesinger {et~al.}(2013)Mesinger, Ferrara, Spiegel, Mesinger, Ferrara,
  \& Spiegel}]{Mesinger2013SignaturesUniverse}
Mesinger, A., Ferrara, A., Spiegel, D.~S., {et~al.} 2013,
  \href{http://dx.doi.org/10.1093/mnras/stt198}{\JournalTitle{Monthly Notices
  of the Royal Astronomical Society}, 431, 621}

\bibitem[{Mesinger {et~al.}(2011)Mesinger, Furlanetto, Cen, Mesinger,
  Furlanetto, \& Cen}]{Mesinger201121cmfast:Signal}
Mesinger, A., Furlanetto, S., Cen, R., {et~al.} 2011,
  \href{http://dx.doi.org/10.1111/j.1365-2966.2010.17731.x}{\JournalTitle{Monthly
  Notices of the Royal Astronomical Society}, 411, 955}

\bibitem[{Mirabel {et~al.}(2011)Mirabel, Dijkstra, Laurent, Loeb, Pritchard,
  Mirabel, Dijkstra, Laurent, Loeb, \& Pritchard}]{Mirabel2011StellarUniverse}
Mirabel, I.~F., Dijkstra, M., Laurent, P., {et~al.} 2011,
  \href{http://dx.doi.org/10.1051/0004-6361/201016357}{\JournalTitle{Astronomy
  {\&} Astrophysics}, 528, A149}

\bibitem[{Ocvirk {et~al.}(2016)Ocvirk, Gillet, Shapiro, Aubert, Iliev,
  Teyssier, Yepes, Choi, Sullivan, Knebe, Gottl{\"{o}}ber, D'Aloisio, Park,
  Hoffman, \& Stranex}]{Ocvirk2016CosmicUniverse}
Ocvirk, P., Gillet, N., Shapiro, P.~R., {et~al.} 2016,
  \href{http://dx.doi.org/10.1093/mnras/stw2036}{\JournalTitle{Monthly Notices
  of the Royal Astronomical Society}, 463, 1462}

\bibitem[{Oh \& Oh(2001)}]{Oh2001ReionizationClusters}
Oh, S.~P., \& Oh, S.~P. 2001,
  \href{http://dx.doi.org/10.1086/320957}{\JournalTitle{The Astrophysical
  Journal}, 553, 499}

\bibitem[{Pawlik {et~al.}(2016)Pawlik, Rahmati, Schaye, Jeon, \&
  Vecchia}]{Pawlik2016TheResults}
Pawlik, A.~H., Rahmati, A., Schaye, J., Jeon, M., \& Vecchia, C.~D. 2016,
  \href{http://dx.doi.org/10.1093/mnras/stw2869}{\JournalTitle{Monthly Notices
  of the Royal Astronomical Society, Volume 466, Issue 1, p.960-973}, 466, 960}

\bibitem[{Petkova {et~al.}(2010)Petkova, Springel, Petkova, \&
  Springel}]{Petkova2010SimulationsFeedback}
Petkova, M., Springel, V., Petkova, M., \& Springel, V. 2010,
  \href{http://dx.doi.org/10.1111/j.1365-2966.2010.17955.x}{\JournalTitle{Monthly
  Notices of the Royal Astronomical Society}, 412, no}

\bibitem[{Pober {et~al.}(2015)Pober, Ali, Parsons, Mcquinn, Aguirre, Bernardi,
  Bradley, Cheng, Deboer, Dexter, Furlanetto, Grobbelaar, Horrell, Jacobs,
  Klima, Kohn, Liu, Macmahon, Maree, Mesinger, Moore, Razavi-Ghods, Stefan,
  Walbrugh, Walker, \& Zheng}]{Pober2015PAPER-64MEDIUM}
Pober, J.~C., Ali, Z.~S., Parsons, A.~R., {et~al.}
  \href{http://dx.doi.org/10.1088/0004-637X/809/1/62}{2015}

\bibitem[{Pritchard {et~al.}(2007)Pritchard, Furlanetto, Pritchard, \&
  Furlanetto}]{Pritchard200721-cmReionization}
Pritchard, J.~R., Furlanetto, S.~R., Pritchard, J.~R., \& Furlanetto, S.~R.
  2007,
  \href{http://dx.doi.org/10.1111/j.1365-2966.2007.11519.x}{\JournalTitle{Monthly
  Notices of the Royal Astronomical Society}, 376, 1680}

\bibitem[{Pritchard \& Loeb(2012)}]{Pritchard2012}
Pritchard, J.~R., \& Loeb, A. 2012,
  \href{http://dx.doi.org/10.1088/0034-4885/75/8/086901}{\JournalTitle{Reports
  on Progress in Physics}, 75, 086901}

\bibitem[{Ricotti {et~al.}(2004)Ricotti, Ostriker, Ricotti, \&
  Ostriker}]{Ricotti2004X-rayMeasurement}
Ricotti, M., Ostriker, J.~P., Ricotti, M., \& Ostriker, J.~P. 2004,
  \href{http://dx.doi.org/10.1111/j.1365-2966.2004.07942.x}{\JournalTitle{Monthly
  Notices of the Royal Astronomical Society}, 352, 547}

\bibitem[{Shapiro {et~al.}(2012)Shapiro, Iliev, Mellema, Ahn, Mao, Friedrich,
  Datta, Park, Komatsu, Fernandez, Koda, Bovill, \& Pen}]{Shapiro2012}
Shapiro, P.~R., Iliev, I.~T., Mellema, G., {et~al.} 2012,
  \href{http://dx.doi.org/10.1063/1.4754363}{in American Institute of Physics
  Conference Series, Vol. 1480, AIP Conference Proceedings, ed. M.~Umemura \&
  K.~Omukai}, 248

\bibitem[{So {et~al.}(2014)So, Norman, Reynolds, \& Wise}]{So2014}
So, G.~C., Norman, M.~L., Reynolds, D.~R., \& Wise, J.~H. 2014,
  \href{http://dx.doi.org/10.1088/0004-637X/789/2/149}{\JournalTitle{The
  Astrophysical Journal}, 789, 149}

\bibitem[{Sobacchi \& Mesinger(2014)}]{Sobacchi2014}
Sobacchi, E., \& Mesinger, A. 2014,
  \href{http://dx.doi.org/10.1093/mnras/stu377}{\JournalTitle{Monthly Notices
  of the Royal Astronomical Society}, 440, 1662}

\bibitem[{Tanaka {et~al.}(2016)Tanaka, O'Leary, Perna, Tanaka, O'Leary, \&
  Perna}]{Tanaka2016TheRadiation}
Tanaka, T.~L., O'Leary, R.~M., Perna, R., {et~al.} 2016,
  \href{http://dx.doi.org/10.1093/mnras/stv2451}{\JournalTitle{Monthly Notices
  of the Royal Astronomical Society}, 455, 2619}

\bibitem[{Trebitsch {et~al.}(2017)Trebitsch, Blaizot, Rosdahl, Devriendt, \&
  Slyz}]{Trebitsch2017FluctuatingGalaxies}
Trebitsch, M., Blaizot, J., Rosdahl, J., Devriendt, J., \& Slyz, A. 2017,
  \href{http://dx.doi.org/10.1093/mnras/stx1060}{\JournalTitle{Monthly Notices
  of the Royal Astronomical Society}, 470, 224}

\bibitem[{van Haarlem {et~al.}(2013)van Haarlem, Wise, Gunst, Heald, McKean,
  Hessels, de~Bruyn, Nijboer, Swinbank, Fallows, Brentjens, Nelles, Beck,
  Falcke, Fender, H{\"{o}}randel, Koopmans, Mann, Miley, R{\"{o}}ttgering,
  Stappers, Wijers, Zaroubi, Akker, Alexov, Anderson, Anderson, van Ardenne,
  Arts, Asgekar, Avruch, Batejat, B{\"{a}}hren, Bell, Bell, van Bemmel,
  Bennema, Bentum, Bernardi, Best, B{\^{i}}rzan, Bonafede, Boonstra, Braun,
  Bregman, Breitling, van~de Brink, Broderick, Broekema, Brouw, Br{\"{u}}ggen,
  Butcher, van Cappellen, Ciardi, Coenen, Conway, Coolen, Corstanje, Damstra,
  Davies, Deller, Dettmar, van Diepen, Dijkstra, Donker, Doorduin, Dromer,
  Drost, van Duin, Eisl{\"{o}}ffel, van Enst, Ferrari, Frieswijk, Gankema,
  Garrett, de~Gasperin, Gerbers, de~Geus, Grie{\ss}meier, Grit, Gruppen,
  Hamaker, Hassall, Hoeft, Holties, Horneffer, van~der Horst, van Houwelingen,
  Huijgen, Iacobelli, Intema, Jackson, Jelic, de~Jong, Juette, Kant,
  Karastergiou, Koers, Kollen, Kondratiev, Kooistra, Koopman, Koster,
  Kuniyoshi, Kramer, Kuper, Lambropoulos, Law, van Leeuwen, Lemaitre, Loose,
  Maat, Macario, Markoff, Masters, McKay-Bukowski, Meijering, Meulman, Mevius,
  Middelberg, Millenaar, Miller-Jones, Mohan, Mol, Morawietz, Morganti,
  Mulcahy, Mulder, Munk, Nieuwenhuis, van Nieuwpoort, Noordam, Norden, Noutsos,
  Offringa, Olofsson, Omar, Orr{\'{u}}, Overeem, Paas, Pandey-Pommier, Pandey,
  Pizzo, Polatidis, Rafferty, Rawlings, Reich, de~Reijer, Reitsma, Renting,
  Riemers, Rol, Romein, Roosjen, Ruiter, Scaife, van~der Schaaf, Scheers,
  Schellart, Schoenmakers, Schoonderbeek, Serylak, Shulevski, Sluman, Smirnov,
  Sobey, Spreeuw, Steinmetz, Sterks, Stiepel, Stuurwold, Tagger, Tang, Tasse,
  Thomas, Thoudam, Toribio, van~der Tol, Usov, van Veelen, van~der Veen, ter
  Veen, Verbiest, Vermeulen, Vermaas, Vocks, Vogt, de~Vos, van~der Wal, van
  Weeren, Weggemans, Weltevrede, White, Wijnholds, Wilhelmsson, Wucknitz,
  Yatawatta, Zarka, Zensus, van Zwieten, \& van Zwieten}]{VanHaarlem2013}
van Haarlem, M.~P., Wise, M.~W., Gunst, A.~W., {et~al.} 2013,
  \href{http://dx.doi.org/10.1051/0004-6361/201220873}{\JournalTitle{Astronomy
  {\&} Astrophysics, Volume 556, id.A2, 53 pp.}, 556}

\bibitem[{Venkatesan {et~al.}(2001)Venkatesan, Giroux, Shull, Venkatesan,
  Giroux, \& Shull}]{Venkatesan2001HeatingBackground}
Venkatesan, A., Giroux, M.~L., Shull, J.~M., {et~al.} 2001,
  \href{http://dx.doi.org/10.1086/323691}{\JournalTitle{The Astrophysical
  Journal}, 563, 1}

\bibitem[{Wouthuysen(1952)}]{Wouthuysen1952OnLine.}
Wouthuysen, S.~A. 1952,
  \href{http://dx.doi.org/10.1086/106661}{\JournalTitle{The Astronomical
  Journal}, 57, 31}

\end{thebibliography}

\end{document}